\documentclass[preprint,12pt]{elsarticle}

\usepackage{amssymb}

\usepackage{lineno}

\usepackage[colorinlistoftodos,prependcaption,textsize=small]{todonotes}
\usepackage{ulem}
\usepackage{soul}

\journal{Carbon}

\begin{document}

\begin{frontmatter}



\title{Water-assisted electronic transport in graphene nanogaps for DNA sequencing}


\author[ift,usp]{Ernane de Freitas Martins}
\author[uff]{Rodrigo G. Amorim}
\author[araraquara]{Gustavo Troiano Feliciano}
\author[suecia]{Ralph Hendrik Scheicher}
\author[ift]{Alexandre Reily Rocha}
\ead{reilya@ift.unesp.br}

\address[ift]{Institute of Theoretical Physics, S\~ao Paulo State University (UNESP), Campus S\~ao Paulo, Brazil}
\address[uff]{Departamento de F\'isica, ICEx, Universidade Federal Fluminense-UFF, Volta Redonda/RJ, Brazil}
\address[araraquara]{Institute of Chemistry, S\~ao Paulo State University (UNESP), Campus Araraquara, Brazil}
\address[suecia]{Department of Physics and Astronomy, Uppsala University, SE-751 20 Uppsala, Sweden}
\fntext[usp]{Current address: Institute of Physics, University of S\~ao Paulo (USP), S\~ao Paulo/SP, Brazil}

\begin{abstract}
Innovative methodologies for reliably and inexpensively sequencing DNA can lead to a new era of personalized medicine. In this work, we performed a theoretical investigation of a nanogap-based all electronic DNA sequencing device. To do so, we used a nitrogen-terminated nanogap on a graphene sheet with the environment fully taken into account. Our investigation is performed using a hybrid methodology combining quantum and classical mechanics coupled to non-equilibrium Green's functions for solving the electron transport across the device. The obtained results show that the DNA nucleotides can be both detected and distinguished in such device, which indicates that it can be used as a DNA sequencing device providing very high sensitivity and selectivity. Furthermore, our results show that water plays a major role in electronic transport in nanoscopic tunneling devices, not only from an electrostatics point of view, but also by providing states that significantly increase the conductance in nanogap-based DNA sequencing devices. 
\end{abstract}


\end{frontmatter}

\section{Introduction}

Personalized DNA sequencing is the next frontier in health care, as it could be used to identify predisposition towards a number of genetic illnesses, and ultimately provide precision treatments \cite{Liang2017,Shendure2017}. In order for this to be fully accomplished, however, the speed and cost of the procedure is yet to drop further \cite{Ansorge2009,Metzker2010,Buermans2016}, and it has become increasingly clear that this can only be achieved via novel sequencing methodologies \cite{Ventra2016,Manuscript2016}.

For this purpose a number of methods have been proposed \cite{dna_sequencing_review}. One such approach is to use nanopores \cite{Lagerqvist2006,Schneider2010,feng2015identification,Feliciano:2015ey,Prasongkit2018} or nanogaps \cite{Prasongkit2013,Amorim2016,Sivaraman2016,Sivaraman2017} acting as sieves for DNA 
segments either measuring electronic or ionic currents as the strands are driven through. The main idea behind an all electronic nanogap sequencing device is to measure conductance differences across the nanogap containing DNA strands. The current is then a signature of the electronic structure of each nucleotide and how its states couple with the nanogap \cite{Amorim2016,Postma2010, Scheicher2012}. 



Amongst possible materials that could be used for such a device, graphene \cite{Novoselov2004} could be an excellent choice as a substrate, since its one-atom thickness could provide single-molecule resolution \cite{Postma2010}. In fact, we have recently shown that functionalizing the edges of graphene nanogaps with nitrogen could enhance the sensitivity and selectivity of graphene-based devices, boosting their potential for sequencing DNA nucleotides \cite{Amorim2016}. Furthermore, we also showed that graphene can be used for detecting specific DNA sequences providing good sensitivity and selectivity \cite{martins2019}.

A key issue in any realistic simulation of bio-sensing devices is that the molecules are embedded in an aqueous solution - typically under physiological conditions. Usually simulations addressing electronic transport in systems containing molecules neglect the effects of the solvent \cite{Prasongkit2013,Amorim2016,amorim2015silicene,Prasongkit2011,DiVentra2013}, or introduce it in the form of a classical potential that only influences the Hamiltonian electrostatically \cite{Feliciano:2015ey,Feliciano2018}. 
A few works have explicitly included the solvent \cite{Rungger2010, Stefano} in theoretical simulations of molecular electronic devices, and in most cases the effect is usually seen as an overall shift of energy levels arising from a chemical gating effect, which are correctly captured by a classical electrostatic effect. Recently we have combined quantum mechanical/classical mechanical methods (QM/MM) with electronic transport to simulate the transport properties of a graphene DNA sequencing device. There we find that the effects of a solvent are important for the correct description of the problem, but explicitly including the electronic states of water does not exert a significant influence compared to a classical external potential added to the quantum mechanical Hamiltonian of the problem. In all cases the devices were conducting.

In this work we show that in the case of tunneling devices, the exact opposite effect occurs. Using a combination of classical and quantum mechanical methods \cite{Feliciano:2015ey,Feliciano2018} in conjunction with the non-equilibrium Green's functions (NEGF) formalism, we show that in a tunneling environment where electrons cross a nanogap, explicitly quantum-mechanically described water molecules play a major role in the overall transport properties of a device to the extent that a significant portion of the current passes through electronic states localized in the water molecules \cite{feng2015identification,Postma2010, Postma2017,DiVentra2016}.




\section{Methodology}

\begin{figure*}[ht]
\center
\includegraphics[width=\textwidth]{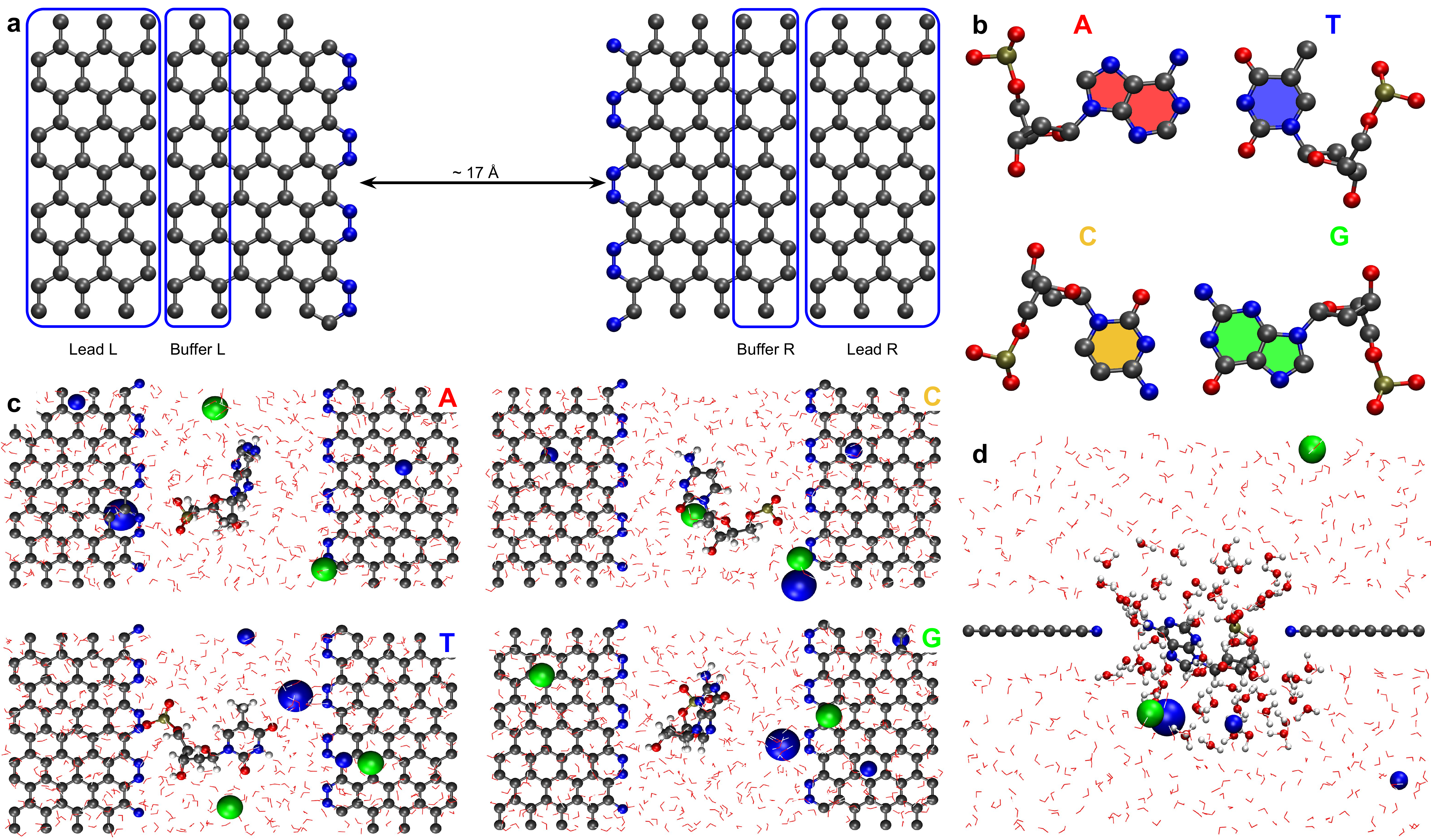}
\caption{a) Setup for electronic transport calculations showing the leads and buffer regions. In this setup we have a scattering problem in which an incoming electron will suffer scattering in the central region being transmitted to the right side of the device. b) Four nucleotides structures (Hydrogen atoms not shown). The four different colors are used to identify the four different nucleotides, and the same color scheme will be used throughout this article. c) Top view of typical snapshots for A, T, C and G nucleotides in the gap. d) Example of QM/MM partition for nucleotide A in setup II, in which the first layer of water is in the QM part. Setup I does not include water in the QM part. The water molecules in ball and stick (line) representation and the large (small) blue spheres - Na atoms - correspond to quantum (classical) partitions. The Cl atoms - green spheres - are always in the classical partitions.}
\label{fig:setup}
\end{figure*}

The device modeled consists of two semi-infinite graphene sheets with nitrogen termination and separated by 17 \AA~ as shown in Figure \ref{fig:setup}a. The four naturally occurring nucleotides are shown in Figure \ref{fig:setup}b: deoxyadenosine monophosphate (A), deoxycytidine monophosphate (C), deoxyguanosine monophosphate (G) and deoxythymidine monophosphate (T). Those nucleotides are placed inside the nanogap-based device, and subsequently filled with water molecules and counter ions with a concentration of 0.2 M. In order to balance the negative charge of the nucleotide, we introduce an imbalance in positive and negative counter ions in such a way that we have 3 Na and 2 Cl atoms in the simulation box. Here we have used isolated nucleotides to simulate the DNA sequencing in line with our previous works \cite{Prasongkit2018,Amorim2016} that have shown little or no effect in the electronic transport due to adjacent bases.



The first step in our approach is to time-evolve the system to realistically simulate dynamical effects. From the structural point of view, as there are no bonds breaking or forming, classical molecular dynamics (MD) yields good results for biological systems \cite{Feliciano:2015ey}. In our simulations the nanogap is kept fixed and only the direction perpendicular to the nanogap is allowed to change (anisotropic NPT simulation). We also apply a harmonic potential in the whole nucleotide to prevent it from moving outside the gap. The systems are simulated in a box with periodic boundary conditions. 

In the employed methodology we first perform a simulation in the isothermal-isobaric ensemble (NPT) using a Berendsen barostat and Nos\'e-Hoover thermostat to equilibrate the density of the system at a given pressure (1 atm). We then extract one snapshot in which the height of the box in the direction perpendicular to the plane of the nanogap ($y$ direction) coincides with the corresponding average. From this initial configuration we perform an isothermal-isochoric ensemble (NVT) production run (same thermostat and restrictions) to generate the structures. Both NPT and NVT simulations were carried out with a 2 fs time step generating 10 ns of dynamics. We have used the AMBER99SB \cite{amber} force field and the SPC \cite{spc1,spc2} model for water, as implemented in GROMACS \cite{gromacs}. 

For each nucleotide, we then select 50 equally separated uncorrelated frames from the NVT production run. Typical snapshots are shown in Figure \ref{fig:setup}c. We then use a QM/MM methodology where each frame is divided \cite{Sanz-Navarro2011,ursuladna} into quantum mechanical (QM) and classical mechanic (MM) regions depending on the setup (see Figure \ref{fig:setup}d for one example). In this work we considered two setups, namely (I) the QM partitions consist of the nitrogenated graphene, the nucleotide and one counter-ion to equilibrate the charge; all water molecules are treated classically, as well as the remaining counter-ions. In the second setup (II) one layer of water is included in the QM partition as shown in Figure \ref{fig:setup}d. A third setup not including solvent effects is presented in the supplementary information for comparison.


For the QM partition we performed density functional theory calculations within the Generalized Gradient Approximation (GGA) as parametrized by Perdew, Burke and Ernzerhof (PBE) \cite{Perdew1996a} with double zeta (DZ) basis set for the carbon atoms of the nanogap and double zeta polarized (DZP) for all remaining atoms. For the atomic core electrons we used Troullier-Martins norm-conserving pseudo-potentials \cite{Troullier1991}. The MM potential is calculated using the same force field, is smoothed out in the electrodes region and is introduced in the DFT calculation of the QM partition. The external potential and DFT calculations were done using the SIESTA package \cite{Soler2002a}. Then, we finally perform the electronic transport calculation \cite{Soler2002a,datta,PhysRevB.73.085414} in these snapshots using a non-equilibrium Green's functions formalism as implemented in Smeagol \cite{PhysRevB.73.085414,Rocha:2005ep}. 
The electronic transport properties were investigated via the NEGF method \cite{datta,PhysRevB.73.085414,Rocha:2005ep, transiesta,  xue2002}. Within that approach, the system is divided into three parts: a scattering region (SR) enclosed by two semi-infinite electrodes on either side \cite{Caroli2001}. The main quantity that can be used to describe the system is the retarded Green's function

\begin{equation}
G^R\left(E,V\right)=\lim_{\eta\rightarrow 0^+}\left[\epsilon S_\mathrm{SR}-H _\mathrm{SR} - \right. \\ \left. {} \Sigma_\mathrm{L}\left(E,V\right) -\Sigma_\mathrm{R}\left(E,V\right)\right]^{-1} ~,
\end{equation}

\noindent where $\epsilon=E+i\eta$, $S_\mathrm{SR}$ and $H_\mathrm{SR}$ are the overlap and Hamiltonian for the scattering region, and $\Sigma\left(E,V\right)_\mathrm{L/R}$ are the so-called self-energies, which include the effects of the electrodes onto the SR. Here the Hamiltonian for the scattering region is taken as the Kohn-Sham Hamiltonian including solvent effects via the external potential. From that, the current can be calculated

\begin{equation}
I\left(V\right)= \frac{2e}{h}\int T\left(E,V\right)\left(f\left(E-\mu_L\right)- \right. \\ \left. {}
f\left(E-\mu_R\right) \right)dE ~,
\end{equation}

\noindent where the quantity

\begin{equation}
T\left(E \right) = Tr \left[\Gamma_{\mathrm R}\left(E,V\right) G^{\mathrm R}\left(E,V\right) \right. \\ \left. {} \Gamma_{\mathrm L}\left(E,V\right) G^{\mathrm A}\left(E,V\right) \right] ~,
\end{equation}

\noindent is the probability that an incoming electron with energy E, will be transmitted across the scattering region. The operators $\Gamma_{L,(R)} \left(E,V\right) = i\left[\Sigma_{L,(R)} - \Sigma_{L,(R)}^\dagger\right] $ define the coupling with the left and right electrodes.

In the limit of zero bias (linear regime), the conductance can be related to the transmission $T\left(E,0\right)$ via the Fisher-Lee relation \cite{FisherLee1981}

\begin{equation}
G = G_0 T\left(E_F\right) ~.
\end{equation}

Finally, by projecting the transmission between any pair of adjacent sites $N$ and $M$, one arrives at the following formula: 

\begin{equation}
i\left(E\right)_{N\rightarrow M} = \\
4\frac{e}{h} \sum_{\tiny 
\begin{array}{c}
n \in N\\ 
m\in M
\end{array}}
\Im
\left[ \left\{ G^{R} \left(E\right) \Gamma_{\mathrm L} G^{A}\left(E\right)\right\}_{mn}  H_{nm} \right] ~,
\label{eq_current}
\end{equation}

\noindent where the above sum runs over all localized atomic orbitals $n$ and $m$ of the basis set belonging to atoms in either site, and for a specific energy $E$. As we are in the limit of zero bias, we considered only right moving components for the current. For further details, we refer to the works of Okabayashi \textit{et al.} \cite{okabayashi2010inelastic} and of Paulsson and Brandbyge \cite{paulsson2007transmission} and Lima \textit{et al.} \cite{PhysRevB.97.165405}.

Finally, given the tunneling nature of the electronic transport here, and consequently its exponential dependence on barrier lengths, we calculate geometric averages of the transmission over snapshots, smoothing out fluctuations resulting in a self-averaging quantity \cite{anderson_localization}.

\section{Results}

\begin{figure}
\centering
\includegraphics[width=0.8\columnwidth]{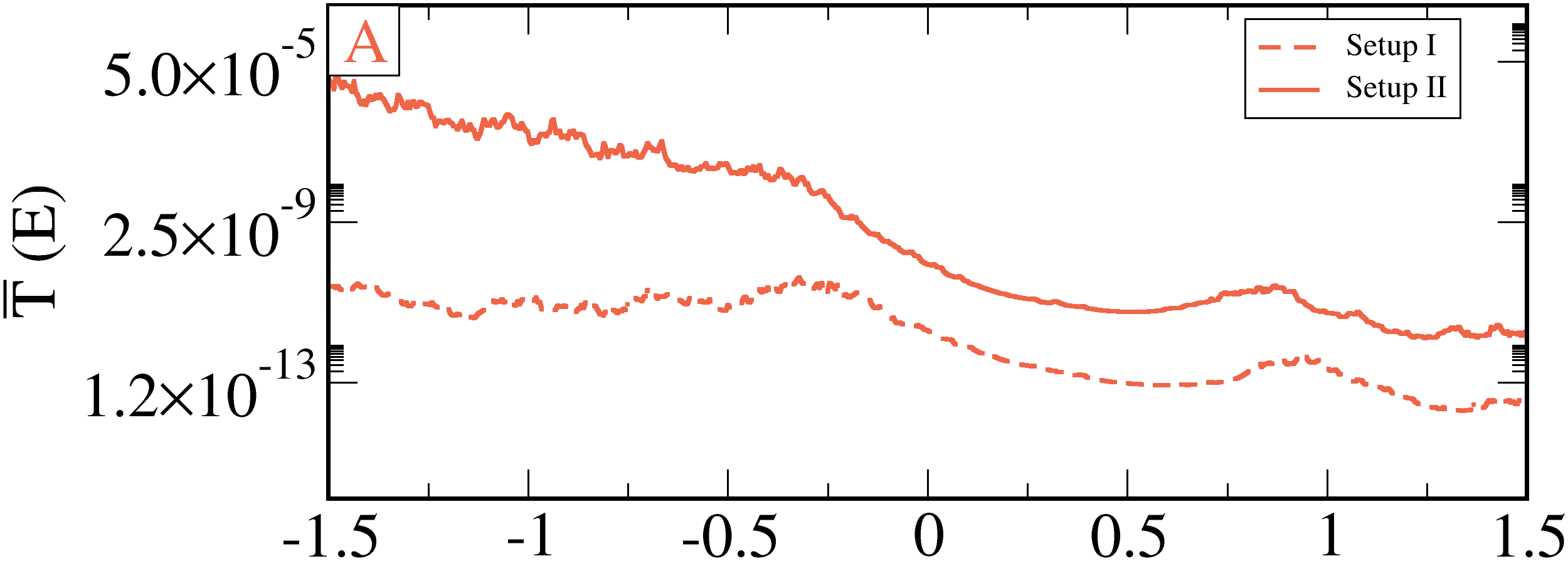}\vskip1ex
\includegraphics[width=0.8\columnwidth]{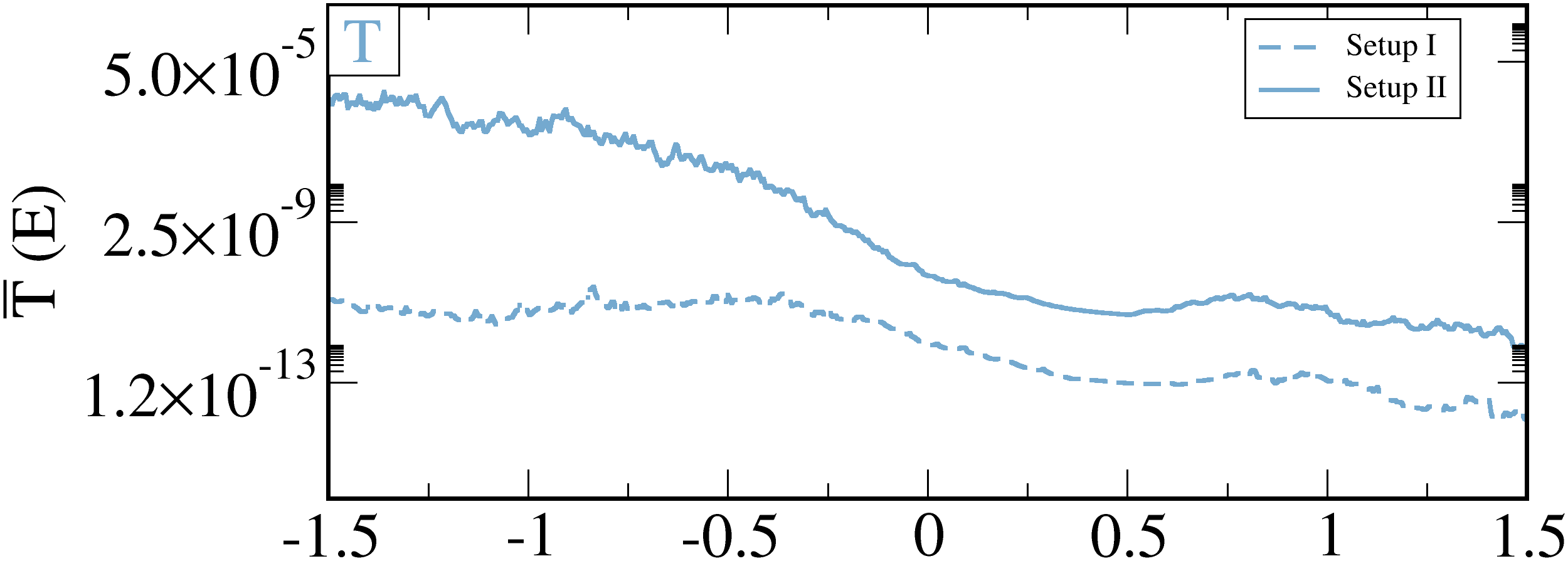}\vskip1ex
\includegraphics[width=0.8\columnwidth]{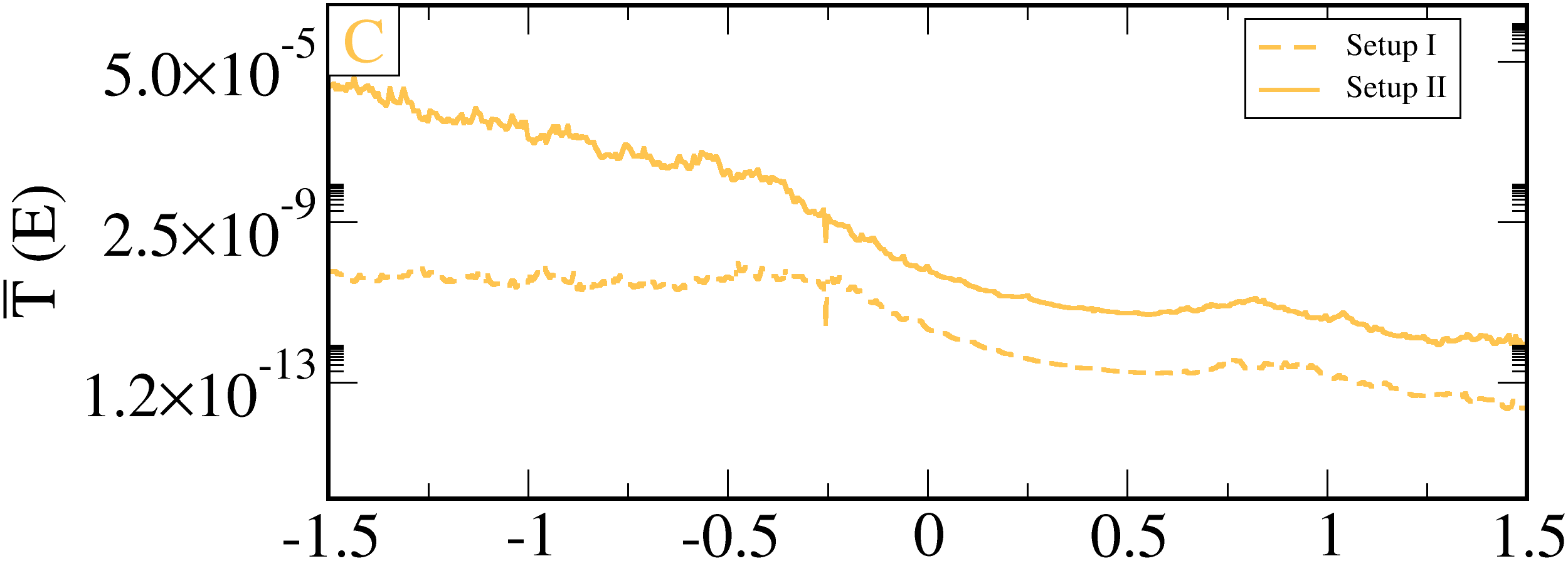}\vskip1ex
\includegraphics[width=0.8\columnwidth]{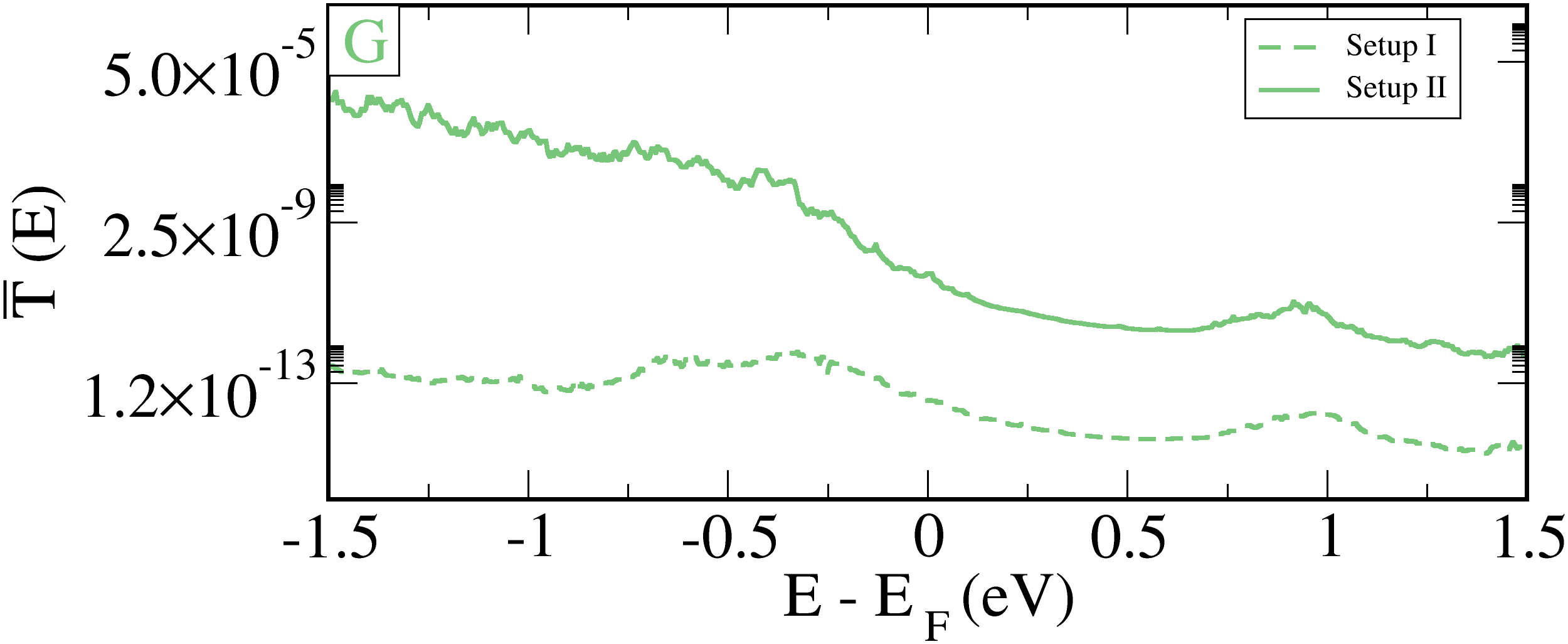}
\caption{Average electronic transmission as function of energy for all nucleotides using different QM partitions. Setup I (dashed line): all water molecules in the MM partition, and setup II (solid line): the same as I including one layer of water molecules in the QM partition.}
\label{fig:all_transp_comp_qm_and_no_qm}
\end{figure}

\begin{figure}
\centering
\includegraphics[width=\textwidth]{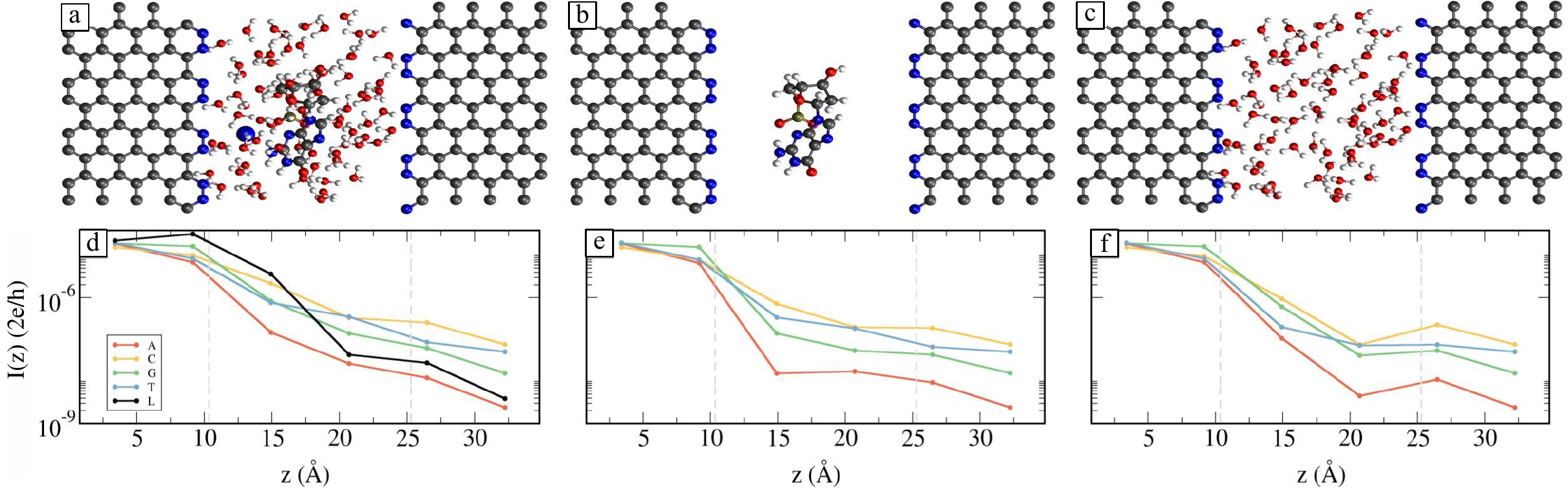}
\caption{Upper panel: on the example of Guanine, we illustrate the different subsystems used to calculate local currents (at zero bias) shown in the lower panels. (a) All atoms treated quantum mechanically, (b) only nitrogenated sheet and the nucleotide treated quantum mechanically and (c) only nitrogenated sheet and water molecules treated quantum mechanically. $z$ component of the local current for the respective subsystem in (d), (e) and (f). The black curve in (d) refers to the system without a nucleotide (reference system).}
\label{graf_correntes_2D_NG}
\end{figure}



Figure \ref{fig:all_transp_comp_qm_and_no_qm} shows the average transmission spectra as a function of energy, for all nucleotides in both setups (I and II), as defined in the Methodology section. Interestingly, for setup I, the magnitude of the transmission at Fermi level is comparable to the system under dry conditions (see supplementary information), but the shape of the curve is significantly different as we are both considering an average behavior as well as including the correct electrostatics from a realistic environment. On the other hand, comparing systems I and II, we note that when the water molecules are considered explicitly (setup II), the average transmission increases for all nucleotides by several orders of magnitude. This result contrasts with previous ones \cite{Rungger2010,Feliciano:2015ey,Feliciano2018} where the inclusion of water molecules only incurs in a gating effect as states coming from water lie far away from the Fermi level \cite{Feliciano2018}.


One way of determining the contribution of water states on these devices is through the analysis of local currents. For this purpose we selected one particular frame with transmission close to the average and calculated the bond current close to the Fermi Level (Equation \ref{eq_current}). Figure \ref{graf_correntes_2D_NG}d shows the total local current projected in the direction of electron flow averaged along the transverse plane. The general profile shown in Figure \ref{graf_correntes_2D_NG}d is a significant decrease as a function of position along the device in the region that comprises the gap, i.e. $10 < z < 25$~\AA. This is expected as we have a tunneling mechanism. We notice that the contribution due to the water molecules is of the same order of magnitude of the local current passing through the nucleotides.



One can also look at the contribution due to different system constituents. Figures \ref{graf_correntes_2D_NG}e and \ref{graf_correntes_2D_NG}f shows this analysis for local currents from one sheet to the other, passing through the nucleotide, and the current that bridges the gap via the water molecules that are treated quantum mechanically.



Figure \ref{fig_correntes_com_cores_T} shows a 3D analysis of the local currents on the example of nucleotide T (see supplementary information for the corresponding graphs for all systems) using different colors to identify the currents of the different parts of the system. The pink arrows represent the current flowing between all the atoms of the sheet, blue arrows represent the current from the sheet to the water molecules and between the water molecules, orange ones represent the current flowing between water molecules and the nucleotide in either direction, and the green ones represent the current flowing between the sheet and the nucleotide in either direction and between the atoms constituting the nucleotide. Since the local current decays exponentially, for clarity we divide the system into 6 regions in the z direction - two inside the gap and two on either electrode - (see supplementary information). The arrow's thickness in Figure \ref{fig_correntes_com_cores_T} is normalized by the largest bond current within each region.

\begin{figure}
\centering
\includegraphics[width=0.8\columnwidth]{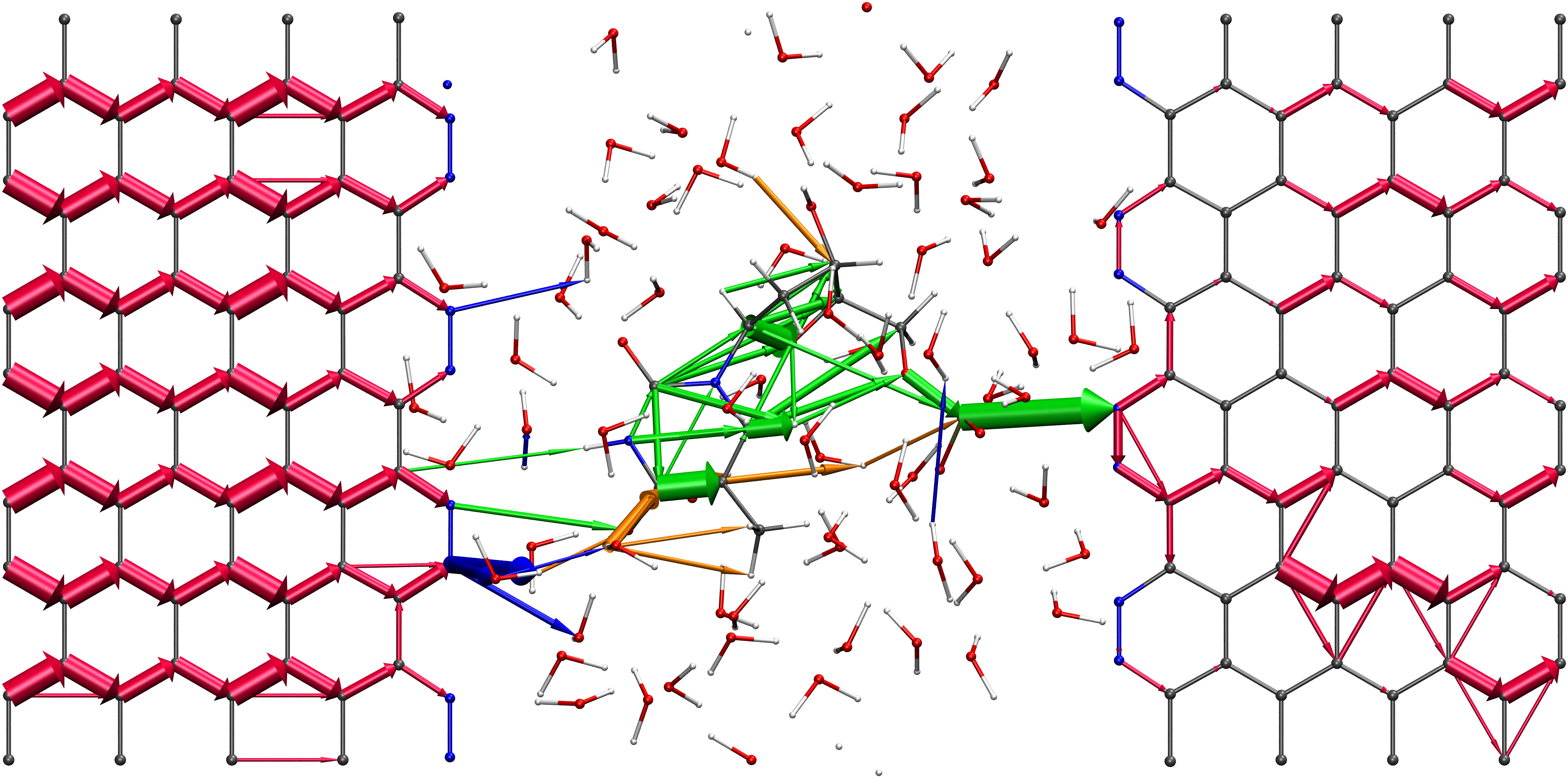}
\includegraphics[width=0.8\columnwidth]{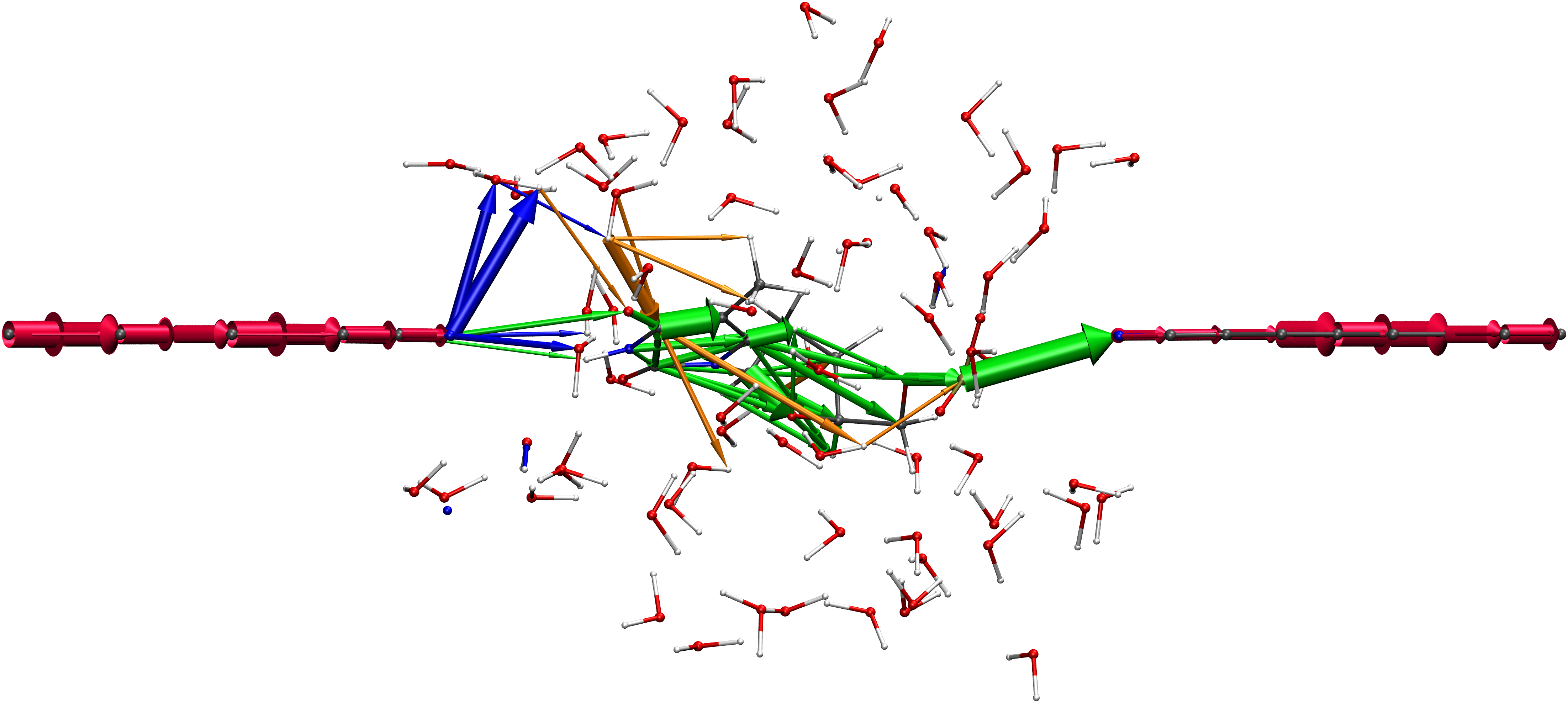}
\caption{Top and side view for the currents on the example of nucleotide T. The arrows' colors identify each subsystem: pink arrows represent the current flowing between all the atoms of the sheet, blue arrows represent the current from the sheet to the water molecules and between the water molecules, orange ones represent the current flowing between water molecules and the nucleotide in either direction, and the green ones represent the current flowing between the sheet and the nucleotide in either direction and between the atoms constituting the nucleotide. A cutoff has been introduced for clarity.}
\label{fig_correntes_com_cores_T}
\end{figure}

As can be observed, the largest currents comes from the graphene sheet (pink arrows) and are initially transmitted from the sheet to the water molecules (blue arrows). Then this current goes from the water molecules to the nucleotide (orange arrows). After that the current flows inside the nucleotide (green arrows) and is transmitted into the right electrode.

The results presented in Figures \ref{graf_correntes_2D_NG} and \ref{fig_correntes_com_cores_T} show why the inclusion of water molecules in the QM part of the system has the effect of increasing the total transmission. As one can observe, the current goes through the water molecules, especially the ones close to the nucleotide and is of particular importance decreasing the barrier between the edges and the core. This is the case because the transmissions at the $E_F$ are not dominated by resonances coming from the nucleotide molecular levels. The alignment of such states at the Fermi level have appeared in a number of simulations (including our own) and can be ascribed to the absence of the solvent and the treatment of the nucleotide as a neutral molecule. Thus, in our case the tail of states from the base and water molecule play similar roles on transport. It is worth mentioning that for the empty gap case the transmittance is null by construction, since there will be no overlap between the states from the left and right electrodes.

\begin{figure}
\centering
\includegraphics[width=0.8\columnwidth]{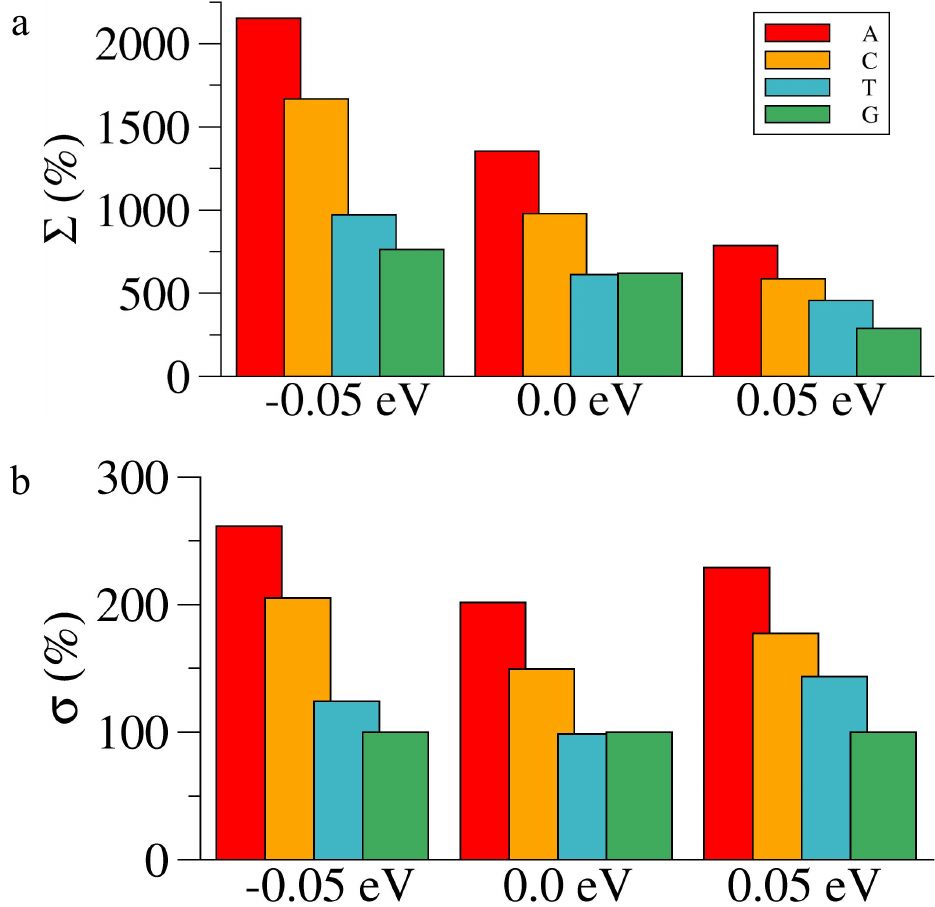}
\caption{(a) Sensitivity for A, C, T and G with respect to the nanogap in liquid without nucleotide, and (b) selectivity for A, C and T with respect to G. Fermi Energy already subtracted.}
\label{fig:selectivity_and_sensitivity}
\end{figure}

Finally, we also performed an analysis of the device's potential sensitivity and selectivity, defined as:


\begin{equation}
\Sigma \ (E') = \frac{\overline{T}_x(E') - \overline{T}_L(E')}{\overline{T}_L(E')} \times 100 \ \%,
\label{eq:sensitivity}
\end{equation}

\noindent and

\begin{equation}
\sigma \ (E') = \frac{\overline{T}_x(E')}{\overline{T}_G(E')} \times 100 \ \%,
\label{eq:selectivity}
\end{equation}

\noindent respectively. Here $\overline{T}_x(E')$ is the average transmission for each nucleotide \hskip1ex (x = A, T, C and G), $\overline{T}_G(E')$ is the average transmission for Guanine (used as our reference) and $\overline{T}_L(E')$ is the average transmission for the nanogap in liquid without the nucleotide (the reference system), at energy $E'=E-E_F$. Figure \ref{fig:selectivity_and_sensitivity} shows the sensitivity and selectivity for each nucleotide in three different energies: $E' = -0.05 \ eV$, $0.0 \ eV$ and $0.05 \ eV$.

By means of sensitivity calculations, we estimate how the presence of any nucleotide compares to the basal conductance. As we can see in Figure \ref{fig:selectivity_and_sensitivity}a, the four nucleotides have extremely high sensitivities for all energies. This means that the presence of one of the bases inside the gap significantly increases the transmission.


At the same time our selectivity calculations, shown in Figure \ref{fig:selectivity_and_sensitivity}b for energies above and below $E_F$ all nucleotides are indeed distinguishable. For $E' = -0.05 \ eV$, for example, Adenine has a transmission around $150 \ \%$ larger than Guanine and the other nucleotides have transmissions between $100 \ \%$ and $20 \ \%$ larger than Guanine at the same energy. For $E' = E_F = 0.0 \ eV$, the nucleotides are also distinguishable, except Thymine, which has virtually the same transmission as the reference. Finally, for $E' = 0.05 \ eV$ we observe that the nucleotides are still distinguishable and Thymine presents the larger selectivity compared to the reference (Guanine). These results for sensitivity and selectivity show that this device for DNA sequencing is highly sensitive and selective for the chosen gate voltages, which are close to Fermi Energy.

\section{Conclusions}

In conclusion we presented a full dynamical treatment to include solvent effects in the investigation of an all-electronic device for DNA sequencing using a graphene sheet containing a nitrogen-terminated nanogap. We used a hybrid methodology that combines quantum mechanics and molecular mechanics (QM/MM) coupled to non-equilibrium Green's functions formalism (NEGF). Our results showed that water plays a fundamental role in the electronic transport properties of this system and that the inclusion of explicit water molecular states increases the total conductance by several orders of magnitude. This result is explained in terms of electron flow along states belonging to water molecules. As the system is away from resonant tunneling the tail of the wave functions of water and base give similar contributions. Finally we also showed that the sensitivity and selectivity of this device is very high, opening up possibilities for further investigations to use it as a DNA sequencing device.

\section{Acknowledgments}

The authors acknowledge financial support from FAPESP (Grant \# FAPESP  2017/02317-2 and 2016/01343-7). A.R.R. acknowledges support from the ICTP-Simons Foundation Associate Scheme. R.G.A. acknowledges financial support from CNPq (2535/2017-1 and 437182/2018-5) and R.H.S. thanks the Swedish Research Council. This study was financed in part by the Coordena\c{c}\~ao de Aperfei\c{c}oamento de Pessoal de N\'{\i}vel Superior - Brasil (CAPES) - Finance Code 001, and  Brazilian Institute of Science and Technology (INCT) in Carbon Nanomaterials, and FAPEMIG and CNPq. The authors also thank the National Laboratory for Scientific Computing (LNCC/MCTI, Brazil) for providing HPC resources of the SDumont supercomputer, which have contributed to the results reported here.

\bibliographystyle{model1-num-names}
\bibliography{references}

\end{document}